\documentclass[aps,prd,twocolumn,superscriptaddress]{revtex4-2} \usepackage{lipsum}
\usepackage{graphicx}  % needed for figures
\usepackage{dcolumn}   % needed for some tables
\usepackage{bm}        % for math
\usepackage{amssymb}   % for math
\usepackage{epsfig}
\usepackage{epstopdf}  
\usepackage{comment}
\usepackage{subfigure}
\usepackage{mdframed}
\usepackage{amsmath}
\usepackage{tikz}
\usetikzlibrary{arrows.meta, positioning}
\tikzset{
  arr/.style={
    ->,
    >=Stealth,
    line width=0.4pt,
    shorten >=1pt,
    shorten <=1pt
  }
}
\usepackage[colorlinks=true]{hyperref}  
\hypersetup{
    bookmarks=true,         % show bookmarks bar?
    unicode=false,          % non-Latin characters 
    pdftoolbar=true,        % show Acrobat
    pdfmenubar=true,        % show Acrobat 
    pdffitwindow=false,     % window fit to page when opened
    pdfstartview={FitH},    % fits the width of the page to the window
    pdftitle={My title},    % title
    pdfauthor={Author},     % author
    pdfsubject={Subject},   % subject of the document
    pdfcreator={Creator},   % creator of the document
    pdfproducer={Producer}, % producer of the document
    pdfkeywords={keyword1} {key2} {key3}, % list of keywords
    pdfnewwindow=true,      % links in new window
    colorlinks=true,       % false: boxed links; true: colored links
    linkcolor=magenta, %red,          % color of internal links (change box color with linkbordercolor)
    citecolor=blue,        % color of links to bibliography
    filecolor=magenta,      % color of file links
    urlcolor=cyan           % color of external links
} 

\newcommand\eea{\end{eqnarray}}
\newcommand\bea{\begin{eqnarray}}

\newcommand{\be}{\begin{equation}}
\newcommand{\ee}{\end{equation}}
\newcommand{\ba}{\begin{align}}
\newcommand{\ea}{\end{align}}
\newcommand{\bg}{\begin{gather}}

\usepackage{graphicx}
\usepackage{dcolumn}
\usepackage{bm}

\begin{document}

%\begin{comment}
\title{Non-invertible symmetries and selection rules for RG flows of coset models}

\author{Valentin Benedetti}
 \email{benedetti@ictp.it}
\affiliation{%
The Abdus Salam International Centre for Theoretical Physics, Strada Costiera 11, Trieste 34151, Italy
}%
\affiliation{%
INFN, Sezione di Trieste, Via Valerio 2, I-34127 Trieste, Italy
} % 
\author{Paul Fendley}
 \email{paul.fendley@physics.ox.ac.uk}
\affiliation{%
The Rudolf Peierls Centre for Theoretical Physics, University of Oxford, Oxford OX1 3PU, UK
} % 
\affiliation{%
All Souls College, University of Oxford
} % 

\author{Javier M. Magan}
 \email{magan@fqa.ub.edu}
\affiliation{%
Departament de F\'isica Qu\`antica i Astrof\'isica, Institut de Ci\`encies del Cosmos\\
Universitat de Barcelona, Mart\'i i Franqu\`es 1, E-08028 Barcelona, Spain
}%
\affiliation{% 
Instituto Balseiro, Centro At\'omico Bariloche, 8400-S.C. de Bariloche, R\'io Negro, Argentina
}%
\date{\today}

\begin{abstract}
We analyze superselection sectors, non-invertible symmetries and selection rules for RG flows triggered via perturbations of a UV two-dimensional conformal field theory (CFT$_2$). To this end we describe a method whose input is the local data, and whose output is the set of submodels of the modular invariant completions. We explain how this output set provides a classification %of possible subcategories 
of superselection sectors (DHR categories and Q-systems) and of topological defect lines, leading to a unified and potentially complete approach to selection rules for RG flows. This method is applied to scenarios in which the UV is a coset or a parafermion model. For these CFT$_2$ we explicitly find all submodels of the diagonal modular invariants. Our results gives selection rules that unify several known facts about such RG flows, while also allowing us to find new aspects.
\end{abstract}
%\end{comment}

\maketitle

%TO compute number of word as of 28/01 use Palabras-totales+Fórmulas-en-texto+16*numero-de-equaciones-312+(210+130)

\textbf{Introduction}: Zamolodchikov's $c$-theorem \cite{Zamolodchikov:1987ti} is a fundamental constraint on renormalization-group flows between two-dimensional conformal field theories (CFT$_2$). Subsequently a variety of tools were developed to analyze flows between rational conformal field theories (RCFT$_2$), the simplest examples being successive ``minimal" models \cite{Belavin:1984vu,Friedan:1984rv}. For example, one can utilize conformal perturbation theory  \cite{Zamolodchikov:1987ti,Ludwig:1987gs}, or exploit symmetries such as supersymmetry \cite{susy3} and/or integrability \cite{CardyTricritical,NonDiagonalRavanini,Fateev2,Fateev3,Zamolodchikov:1991vg}.
In more recent years, further fundamental information has been obtained by utilizing non-invertible symmetries \cite{Petkova:2000ip,Frohlich:2006ch} and their description using fusion categories, in both unitary \cite{Chang:2018iay,Damia:2024xju,Ambrosino:2025yug,Arias-Tamargo:2025atr,Gaberdiel:2026sfg,Ambrosino:2026umb}
and non-unitary models 
\cite{Jacobsen:2023isq,Nakayama:2024msv,Katsevich:2024jgq,Delouche:2024yuo}.

Understanding the classification and analysis of \emph{all} possible selection rules is an obvious important task and valuable tool. An RCFT$_2$ contains a wealth of data beyond the symmetries, including that of superselection sectors. The Doplicher, Haag and Roberts (DHR) approach \cite{DHR1,DHR2} provides a solid ground to explore this data. This approach was originally designed within local quantum field theory \cite{Haag:1963dh,Araki:1981gy,Haag:1996hvx,haag2012local}, and is based on the category of localized endomorphisms of the observable algebra \footnote{To be precise, DHR superselection sectors \cite{DHR1,DHR2} of a given model are inequivalent Hilbert-space representations characterized by being unitarily equivalent to the vacuum outside a given finite region of trivial topology (in the present two-dimensional case, outside a given spatial segment). Intuitively, such sectors are created by applying certain local operators (vertex-type in RCFT$_2$) to the vacuum. By the usual argument \cite{DHR1,DHR2}, DHR sectors are unitarily equivalent to endomorphisms of the observable algebra, and the latter form a $C^*$ category with direct sums and subobjects \cite{haag2012local}.}. A key aspect is that the CFT$_2$ data is sufficient to understand precisely how the DHR category changes with any relevant perturbation \cite{Benedetti:2024utz}. While some perturbations destroy the CFT$_2$ superselection sectors entirely, others do not.
A putative CFT$_2$ infrared (IR) fixed point must then possess any remaining structure, strongly constraining the flow. This perspective nicely connects  with the approach to generalized order parameters based on Haag duality violations \cite{Casini:2019kex,Casini:2020rgj,Review,Benedetti:2022zbb,Benedetti:2024dku,Casini:2025lfn}, allowing the computation of relative Jones indices \cite{Jones1983,Kosaki1986ExtensionOJ,L11}, as we review below.

Given this context, our first goal is to describe a unified approach able to capture all selection rules, with the features that it is systematic, non-perturbative, and universally valid across CFT$_2$. Our second goal is to demonstrate the power of this method within  two prominent families of RCFT$_2$ \cite{Goddard:1972iy, Goddard:1986bp}. This entails a new classification of symmetries, superselection sectors and $Q$-systems across a broad class of $c \geq 1$ theories, greatly extending the classification for minimal models \cite{Kawahigashi:2002px,Kawahigashi:2003gi,Benedetti:2024utz}. It will also lead to a non-perturbative explanation of observed RG flows for integrable deformations \cite{Zamolodchikov:1987ti,Zamolodchikov:1991vg,Fateev3}.

Our framework makes transparent the intricate connections between apparently disparate modern techniques, such as those based on symmetry and DHR categories, on standard CFT data, on the analysis of Haag duality violations, and on the classification of CFT$_2$'s. It further connects the analysis of selection rules with the classification of $Q$-systems \cite{Longo:1994zza,Longo:1994xe,LongoKawaMuger,bischoff2015tensor}, which are certain algebraic structures in the DHR category \footnote{In precise mathematical terms, a $Q$-system in a C$^*$-category of endomorphisms of a type $III$ algebra $\mathcal{N}$ (such as a the DHR category) is a triple $\lbrace\rho,\omega,x\rbrace$, where $\rho$ is an object in the category (a unital endomorphism of the algebra $\mathcal{N}$), $w\in\mathcal{N}$ is an isometry intertwining the identity representation with $\rho$, $x\in\mathcal{N}$  is an isometry intertwining the $\rho$ representation with $\rho^2$, and $w,x$ satisfy particular algebraic relations guaranteeing that $\rho$ is the dual canonical
endomorphism associated with an algebraic extension $\mathcal{N}\subset\mathcal{M}$, see \cite{bischoff2015tensor} for a brief but complete exposition.},  whose importance in this context relies in the fact  they control local extensions of chiral algebras \cite{bischoff2015tensor}.

\textbf{Beyond modular invariance and selection rules:}
Modular invariance is an extremely valuable tool in both classifying and analyzing the properties of CFT$_2$ \cite{Cardy:1986ie}, and it is typically required in axiomatic approaches \cite{Segal:2002ei}. However, CFTs have much more structure not apparent in modular-invariant partition functions. They can possess fermionic or parafermionic chiral operators \cite{Moore:1988qv}, and be defined with more general open or twisted boundary conditions \cite{Cardy:1986gw,Cardy:1989ir}. 
Moreover, modular invariance does not follow naturally from the basic axioms of local quantum field theory \cite{haag2012local}. In the torus, modular transformations can be decomposed into $T$-transformations, acting on the torus parameter $\tau$ as $\tau\to\tau+1$, and $S$-transformations, implementing $\tau\to -1/\tau$. While invariance under $T$ is necessary for operators to commute when space-like separated \footnote{One can define a more general \emph{graded} notion of $T$ duality allowing for fermionic primaries and anticommuting operators}, invariance under $S$ is not \cite{Alvarez:1994dn,Rehren:2000ti}. In this vein, $S$-invariance was conjectured to be equivalent to the absence of superselection sectors \cite{Rehren:2000ti}. A recent proof \cite{Benedetti:2024dku} (see also \cite{LongoKawaMuger}) connects it to the absence of Haag-duality violations, and therefore to the notion of completeness of the local-operator algebras \cite{Review}.

Given these observations, a basic question concerns classifying all local CFT$_2$, including those not $S$-invariant.  A rigorous method to achieve this goal is the classification of $Q$-systems of chiral algebras and their extensions \cite{Longo:1994zza,Longo:1994xe,LongoKawaMuger,bischoff2015tensor,Kawahigashi:2002px,Kawahigashi:2003gi}. A simpler method, connecting with standard bootstrap techniques, was explored in \cite{Benedetti:2024utz}.  It starts with a modular-invariant CFT$_2$ $\mathcal{C}$, characterized by a set of families $\left[ \phi_i\right]$ obeying some fusion rules. A local submodel $\mathcal{T}$ is then defined from a set of families closed under fusion. For RCFT$_2$ this procedure is finite, and all submodels $\mathcal{T}\subset \mathcal{C}$ can be found. When $\mathcal{C}$ is diagonal (the partition function is of the form $\sum_j \chi_j(\tau)\overline{\chi}_j(\tau)$), we can use the chiral fusion rules $\phi_a\phi_b = \sum_c N_{ab}^c \phi_c$, where the fusion matrices follow from the Verlinde formula \cite{VerlindeFusion}
\be
N^{k}_{ij}=\sum_n\frac{S_{in} S_{jn} \bar{S}_{nk}}{S_{0n}}\;,\label{verlinde-fusion}
\ee
with the modular $S$ matrix arising from the $S$-transformation of characters $\chi_i(-1/\tau)=\sum_j\,S_{ij}\chi_j(\tau)$.

How does this classification relate to selection rules? To answer this, consider a UV CFT$_2$ perturbed with a relevant $\phi_{UV}$. From an algebraic standpoint, the key is to view the flow as one in the smallest submodel $\mathcal{T}_{UV}[\phi_{UV}]$ containing $\phi_{UV}$. The reason is that such perturbation can at most affect other families in $\mathcal{T}_{UV}[\phi_{UV}]$. Then, the remaining operator content behaves as a spectator \cite{Benedetti:2024utz}, leading to the following algebraic selection rule \footnote{Mathematically, one can state this selection rule as the preservation of the set of possible $Q$-systems for $\mathcal{T}_{UV}[\phi_{UV}]$, see the introduction section for the definition of $Q$-systems.
}:

\vspace{0.15cm}
\noindent \textit{The structure of extensions $\mathcal{T}\supset \mathcal{T}_{UV}[\phi_{UV}]$ %of $\mathcal{T}_{UV}[\phi_{UV}]$ 
is preserved along the RG flow.}
\vspace{0.15cm}

\textbf{Superselection sectors and topological defects:}  We here explain how selection rules coming from superselection sector and symmetry categories descend from the previous local algebraic selection rule. Let's begin with DHR sectors  \cite{DHR1,DHR2}. These appear when the local algebras cannot produce all localized finite-energy states from the vacuum. For example, a modular-invariant CFT$_2$ $\mathcal{C}$ has no DHR sectors, as its local operators generate the entire Hilbert space \cite{LongoKawaMuger,Benedetti:2024dku}; it is ``complete'' \cite{Review,Benedetti:2024dku}. Conversely, a submodel $\mathcal{T}\subset\mathcal{C}$ does not contain certain charge-creating local operators that exist in  $\mathcal{C}$ and not in $\mathcal{T}$; e.g., if  $\mathcal{C}$ is the Ising model and $\mathcal{T}$ is the submodel formed by the stress tensor, then the $\sigma$ and $\epsilon$ fields are in $\mathcal{C}$ but not in $\mathcal{T}$. These  operators create local charges in $\mathcal{T}$, namely they create DHR sectors. For CFT$_2$, these are unitarily equivalent to localized endomorphisms of $\mathcal{T}$, and form a braided category \cite{Buchholz,Fredenhagen:1988fj,LongoKawaMuger,RehrenChiral,haag2012local}.

We now describe a practical method to find the DHR category associated with any RCFT$_2$ $\mathcal{T}$, i.e., the category of localized and transportable endomorphisms of $\mathcal{T}$ \cite{haag2012local}. Following \cite{LongoKawaMuger, Casini:2025lfn}, an insightful probe of this category is the  Jones index \cite{Jones83,Kosaki1986ExtensionOJ,L11} of the two-interval inclusion of algebras in $\mathcal{T}$. The Jones index associated with a particular inclusion of algebras $\mathcal{N}\subset\mathcal{M}$ is a rigorous measure of the relative size between the two, and, more importantly, it is controlled by the category of endomorphisms of $\mathcal{N}$ arising in the particular embedding in $\mathcal{M}$ \cite{LongoNets}. For the present case, consider two intervals $I_1$ and $I_2$, and their complementary region $I_c\equiv (I_1\cup I_2)'$. Call $\mathcal{A}_{\mathcal{T}}(I)$ to the algebra of operators in $\mathcal{T}$ that are fully localized in $I$. The two-interval inclusion is then
\be\label{inca}
\mathcal{A}_{\mathcal{T}}(I_1\cup I_2)\subseteq \mathcal{A}_{\mathcal{T}}(I_c)'\;.
\ee
The associated global Jones index $\mu_{\mathcal{T}}$ is independent of the choice of intervals, and is the square of the total quantum dimension of the DHR category \cite{LongoKawaMuger}. Furthermore, the Jones index $\lambda$ of an inclusion $\mathcal{T}\subset\mathcal{T}'$ is related to the global indices of each model as $\mu_{\mathcal{T}}=\lambda^2\, \mu_{\mathcal{T}'}$ \cite{LongoKawaMuger}, so computing global indices implies computing relative indices as well \footnote{This formula is only valid in $d=2$. For $d>2$ the formula is modified to $\mu_{\mathcal{T}}=\lambda\, \mu_{\mathcal{T}'}$, see Ref. \cite{Casini:2025lfn}.}. An explicit formula for $\mu_{\mathcal{T}}$ was found in \cite{Benedetti:2024dku}, see also \cite{Kawahigashi:2004rf}. It only uses chiral data (the $S$-matrix), and the multiplicities $Z_{ij}$ of the conformal families appearing in $\mathcal{T}$ whose left- and right-moving parts are labeled by $i$ and $j$ respectively. More precisely
\be 
\mu_{\mathcal{T}}= \left(\frac{\sum_{i}d_{i}^2}{\sum_{j}\sum_{k}Z_{jk}d_j d_{k}}\right)^2\,, \quad d_{i}=\frac{{S}_{0i}}{{S}_{00}} \;,\label{index-global}
\ee
where $d_{i}$ are the quantum dimensions \footnote{Although this expression might seem ``chiral'', it is not. The sum $\sum_{i}d_{i}^2$ is the total quantum dimension of the category. It will be the same for a different (non-diagonal) completion, for which it will take a more general form $\sum_{jk}Z_{jk}d_j d_{k}$, with $Z_{jk}$ the multiplicities defining the completion. Then, it is clear from Eq. \eqref{index-global}, that complete models have $\mu_{\mathcal{C}}=1$}. Then, having classified all submodels $\mathcal{T}$, we can compute $\mu_{\mathcal{T}}$ for each $\mathcal{T}$, suggesting appropriate ansatz for the categories. These ansatz are verified by unravelling the fusion rules in the new primary classes, defined not under the chiral algebra, but under the $\mathcal{T}$ algebra. From this perspective, the previous algebraic selection rule is reframed as \footnote{This selection rule also follows from ``transportability of the DHR sectors'', see \cite{Casini:2025lfn} for a precise definition of transportability, and appropriate references to previous literature.}

\vspace{0.15cm}
\noindent \textit{The DHR category of $\mathcal{T}_{UV}[\phi_{UV}]$ is preserved along the RG flow and reproduced in IR by a submodel $\mathcal{T}_{IR}[\phi_{IR}]$.} 
\vspace{0.15cm}

We continue with the symmetry category of a modular invariant $\mathcal{C}$, with respect to the chiral algebra,  
and seek to derive its fusion ring. The symmetries here are generated by Topological Defect Lines  (TDLs), namely line operators that commute with all chiral-algebra generators \cite{Petkova:2000ip,Frohlich:2006ch}. These are then linear combinations of projectors $P_{ij}\equiv\vert [ij]\rangle\langle [ij]\vert$  onto the Verma modules $V_i\otimes \bar{V}_j$ defining $\mathcal{C}$ (or more general operators of the form $\vert [ij]a\rangle\langle [ij]b\vert$, with $a,b=1,\cdots,Z_{ij}$ if $Z_{ij}>1$), i.e.\ projections into the different superselection sectors. For diagonal models $P_{ij}\Rightarrow \delta_{ij}P_i$, and the standard basis of Verlinde lines follows from the modular $S$ matrix as $\mathcal{L}_{i}\equiv \sum_{j}(S_{ij}/S_{0j})P_{j}$ (or equivalently $P_{j}=S_{0j}\sum_{i}S_{ji}^{*} \mathcal{L}_{i}$), obeying the same fusion rules as DHR sectors $
\mathcal{L}_{i}\times\mathcal{L}_{j}=\sum_{k}N^{k}_{ij}\mathcal{L}_{k}$ (albeit the categories might differ). We can extend these observations \footnote{This generalization can refer to a non-diagonal completion of a chiral algebra, but also to the relation between a completion and a submodel larger than the chiral algebra, or a combination of both scenarios thereof.} to find the symmetries of a modular invariant $\mathcal{C}$ with respect to a submodel $\mathcal{T}$. The DHR category associated to $\mathcal{T}$ allows $\mathcal{C}$ to be decomposed as a direct sum of sectors of $\mathcal{T}$, in the same way that a RCFT$_2$ is decomposed as a sum of chiral sectors. The symmetry algebra is then isomorphic (up to multiplicities) to the algebra of projectors into those new enlarged sectors. These projectors are the ``TDLs'' with respect to an enlarged algebra, namely that of the submodel $\mathcal{T}$, and are defined as the lines commuting with $\mathcal{T}$. More concretely, the Hilbert space $\mathcal{H}_{\mathcal{C}}$ of $\mathcal{C}$ decomposes as $\mathcal{H}_{\mathcal{C}}= \oplus_r\,N_r\pi_r (\mathcal{T})$ in terms of inequivalent DHR representations of $\mathcal{T}$ and their multiplicities $N_r$. The projectors $p_r$ into each block $\pi_r (\mathcal{T})$ (defined as $p_r\vert r'\rangle\equiv \delta_{rr'}\vert r\rangle$, with $\vert r\rangle\in \pi_r (\mathcal{T})$), commute with $\mathcal{T}$ and are therefore TDLs with respect to it.

For diagonal $\mathcal{C}$ we have simplifications. Since here the TDL fusion is isomorphic to the DHR fusion, the closed TDL rings in $\mathcal{C}$ are in one-to-one correspondence with the closed submodels $\mathcal{T}\subset \mathcal{C}$. Also, if $\mathcal{T}$ has a certain  DHR fusion, there is an isomorphic TDL ring $\mathfrak{F}$ commuting with $\mathcal{T}$. Then, $\mathcal{T}$ can be defined by specifying the fusion ring $\mathfrak{F}$ commuting with all operators in $\mathcal{T}$ as in \cite{Shao:2025mfj,Zhang:2025bsm}. In fact,  we can use the TDLs in  $\mathfrak{F}$ to build a projector onto $\mathcal{T}$ \footnote{We can also use the TDL forming  $\mathfrak{F}$ to build projectors into any DHR superselection sector associated to $\mathcal{T}$.} (see \cite{Saura-Bastida:2024yye,Molina-Vilaplana:2025vzt,AliAhmad:2025bnd,Benini:2025lav,Zhang:2025bsm} for related discussions). In terms of the projectors onto the Verma modules that define $\mathcal{T}$ we have $P_\mathcal{T}\equiv\sum_{j\in \mathcal{T
}}P_{j}$. For the present discussion, it can be nicely written with Verlinde lines in $\mathfrak{F}$ as
\be
P_\mathcal{T}= \frac{1}{\sqrt{\mu_{\mathcal{T}}}}\,\sum_{i\in\mathfrak{F}}d_i \mathcal{L}_i\;.
\ee
Since the fusion rings of TDL's and DHR associated with $\mathcal{T}$ are isomorphic, for diagonal models we obtain
\be 
\mu_{\mathcal{T}}= \Big(\sum_{i\in\mathfrak{F}}d_{i}^2\Big)^2 \,.\label{index-verlinde}
\ee
This result shows the global Jones index $\mu_{\mathcal{T}}$, computed solely with CFT data \eqref{index-global}, provides a sharp requirement for the symmetry category.
Concerning selection rules, this discussion shows the algebraic selection rule implies

\vspace{0.15cm}
\noindent \textit{The fusion ring  $\mathfrak{F}_{UV}[\phi_{UV}]$  is preserved along the flow and reproduced by $\mathfrak{F}_{IR}[\phi_{IR}]$ for some $\phi_{IR}$.} 
\vspace{0.1cm}

\noindent 
A consequence is that all TDL commuting with $\phi_{UV}$ are reproduced in the IR, as originally discussed in \cite{Chang:2018iay}. 

We conclude that all information about DHR categories, $Q$-systems, symmetry fusion rings, and selection rules ultimately stem from local CFT data at the UV, in particular from the classification of local RCFT$_2$. Below we verify all these features in a large set of of models.

\textbf{RG flows between cosets:} We use these ideas to study RG flows between coset models,  a construction pioneered by Goddard, Kent and Olive
\cite{Gko1,Gko2}; for a review see \cite{Goddard:1986bp}. One starts with a Wess-Zumino-Witten (WZW) RCFT$_2$ $G_k$,  with Lie-group symmetry $G$, and with $k$ the integer coefficient of the WZW term in the action \cite{Wess:1971yu,Witten:1983tw}.  
The coset model $G_k/H_{k'}$ is then obtained by gauging a subgroup $H$ of $G$. The notation reflects the intuition that this gauging is tantamount to ``modding out" $G_k$ by another WZW model $H_{k'}$, where $k'$ is fixed by how $H$ acts on $G_k$. We focus on two specific families in this paper. The first is comprised of the diagonal $\mathfrak{su}(2)$ cosets 
\be 
\text{D}(k,l)= \frac{\mathfrak {su}(2)_k \times \mathfrak{su}(2)_l}{\mathfrak{su}(2)_{k+l} }\,.\label{coset-gko}
\ee
These models have been extensively studied, admitting a lattice analog \cite{Date:1986ju,Date:1987wy} %via Restricted-Solid-On-Solid models \cite{Andrews:1984af},
and a Coulomb gas interpretation \cite{DiFrancescoCoulombGas}. The modular-invariant partition functions have been classified \cite{Kmq,Ravaninigko1,Ravaninigko2}, taking into account the extended algebras for $c>1$ \cite{Bagger:1987kv}. Moreover, the models D$(k,1)$ correspond to the minimal models \cite{Belavin:1984vu,Friedan:1983xq}, while D$(k,2)$ correspond to the $\mathcal{N}=(1,1)$  superconformal minimal models \cite{Susy4,Susy1}.

For arbitary $k$ and $l$, the central charge is 
\be
c_{\text{D}(k,l)}=\frac{3kl(k+l+4)}{(k+2)(l+2)(k+l+2)}\;.
\ee
and the chiral fields are labeled by three parameters $(\alpha,\beta,\gamma)$, associated with the affine Lie algebras in \eqref{coset-gko}:
\[
\alpha=0,1,\dots,k ,\,\,\, \beta=0,1\dots,l,\,\,\,  \gamma=0,1\dots,k+l\;,
\]
with $\gamma$ appearing in $\alpha \beta$ mod $2$. We focus on diagonal completions in order to find submodels using the fusion rules (\ref{verlinde-fusion}). The $S$ matrix of the coset factorizes into the corresponding WZW models via $S^{\mathfrak{su}(2)_k}_{\alpha\alpha'}S^{\mathfrak{su}(2)_l}_{\beta\beta'}\overline{S}^{\mathfrak{su}(2)_{k+l}}_{\gamma\gamma'}$ (up to a normalization \cite{DiFrancesco:1997nk}), with
\be 
S_{\alpha\alpha'}^{\mathfrak{su}(2)_k}=\sqrt{\frac{2}{k+2}}\sin\left[\frac{\pi(\alpha+1)(\alpha'+1)}{k+2}\right]\,.  \label{smat-wzw}
\ee
We note the model includes a simple current enforcing the indentification $(k-\alpha,l-\beta,k+l-\gamma)\leftrightarrow(\alpha,\beta,\gamma)$. For $k$ and $l$ both even, the model requires fixed-point resolution of this $\mathbb{Z}_2$ symmetry, as addressed in \cite{Schellekens:1989uf,Fuchs:1995tq}.
%, but we will not consider these cases below.

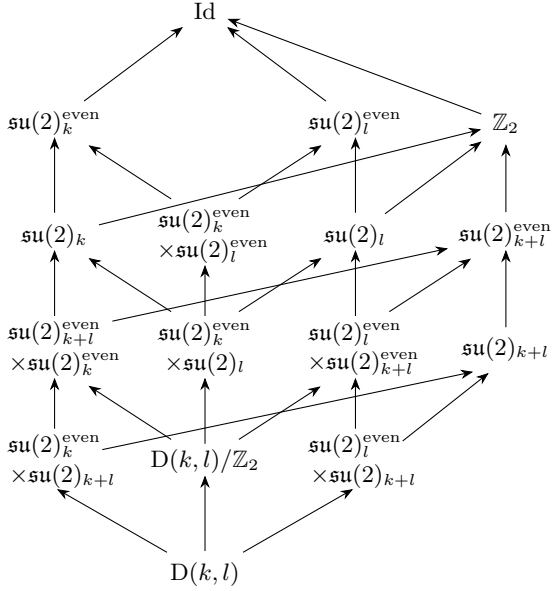
\begin{figure}[ht]
\centering
\begin{tikzpicture}[
    node distance=2.2cm and 3.0cm,
    every node/.style={font=\small},
    -/.style={->, thick}
]
%\node (topL) at (-6,6) {$k\ \text{odd}\quad l\ \text{odd}\quad k+l\ \text{even}$};
\node (Id)   at (-4,6) {$\text{Id}$};
%------------------------------------------------
\node (suke) at (-6,4.5) {$\mathfrak{su}(2)_k^{\text{even}}$};
\node (sule) at ( -2,4.5) {$\mathfrak{su}(2)_l^{\text{even}}$};
\node (Z2)  at (0,4.5) {$\mathbb{Z}_2$};
%------------------------------------------------
\node (suk)  at (-6,3) {$\mathfrak{su}(2)_k$};
\node (prod1) at (-4,3.2) {$\mathfrak{su}(2)_k^{\text{even}}$};
\node (prod12) at (-4,2.8) {$\,\,\,\,\times \mathfrak{su}(2)_l^{\text{even}}$};
\node (sul)  at ( -2,3) {$\mathfrak{su}(2)_l$};
\node (sukle) at ( 0,3) {$\mathfrak{su}(2)_{k+l}^{\text{even}}$};
%------------------------------------------------
\node (prod2L) at (-6,1.7) {$\mathfrak{su}(2)_{k+l}^{\text{even}}$};
\node (prod2L2) at (-6,1.3) {$\,\,\,\,\,\times \mathfrak{su}(2)_k^{\text{even}}$};
\node (prod2M) at (-4,1.7) {$\mathfrak{su}(2)_k^{\text{even}}$};
\node (prod2M2) at (-4,1.3) {$\times \mathfrak{su}(2)_l$};
\node (prod2R) at ( -2,1.7) {$\mathfrak{su}(2)_{l}^{\text{even}}$};
\node (prod2R2) at ( -2,1.3) {$\,\,\,\,\times \mathfrak{su}(2)_{k+l}^{\text{even}}$};
\node (sukl)  at ( 0,1.5) {$\mathfrak{su}(2)_{k+l}$};
%------------------------------------------------
\node (prod3L) at (-6,0.2) {$\mathfrak{su}(2)_{k}^{\text{even}}$};
\node (prod3L2) at (-6,-0.2) {$\,\,\,\,\times \mathfrak{su}(2)_{k+l}$};
\node (coset)  at ( -4,0) {$\text{D}(k,l)/{\mathbb{Z}_2}$};
\node (prod3R) at ( -2,0.2) {$\mathfrak{su}(2)_{l}^{\text{even}}$};
\node (prod3R2) at ( -2,-0.2) {$\,\,\,\,\times \mathfrak{su}(2)_{k+l}$};
%------------------------------------------------
\node (bottom) at (-4,-1.5) {$\text{D}(k,l)$};
%------------------------------------------------
\draw[arr] (suke) -- (Id);
\draw[arr] (sule) -- (Id);
\draw[arr] (Z2) -- (Id);
%------------------------------------------------
\draw[arr] (suk) -- (-6,4.35);
\draw[arr] (sul) -- (-2,4.35);
\draw[arr] (prod1) -- (sule);
\draw[arr] (prod1) -- (suke);
\draw[arr] (sukle) -- (Z2);
\draw[arr] (sul) -- (Z2);
\draw[arr] (suk) -- (Z2);
%------------------------------------------------
\draw[arr] (-6,1.9) -- (-6,2.85);
\draw[arr] (-4,1.9)-- (-4,2.65);
\draw[arr] (-2,1.9) -- (-2,2.85);
\draw[arr] (0,1.75) -- (0,2.85);
\draw[arr] (prod2L) -- (sukle);
\draw[arr] (prod2M) -- (suk);
\draw[arr] (prod2M) -- (sul);
\draw[arr] (prod2R) -- (sukle);
%------------------------------------------------
\draw[arr] (-6,0.4)  -- (-6,1.15) ;
\draw[arr] (-2,0.4) -- (-2,1.15) ;
\draw[arr] (-4,0.2)  -- (-4,1.15) ;
\draw[arr] (coset) -- (prod2L2);
\draw[arr] (coset) -- (prod2R2);
\draw[arr] (-1.4,0.25) -- (-0.2,1.2);
\draw[arr] (-5.4,0.15) -- (-0.4,1.2);
%------------------------------------------------
\draw[arr] (bottom) -- (-6,-0.35);
\draw[arr] (bottom) -- (-4,-0.15);
\draw[arr] (bottom) -- (-2,-0.35);
\end{tikzpicture}
\caption{Classification of submodels of the coset model D($k,l$) \eqref{coset-gko} for $k,l$ odd, with each denoted by its DHR  category.}
\label{figure-gko-letter}
\end{figure}

Applying our procedure, we find that the diagonal modular invariant of $\text{D}(k,l)$ has sixteen submodels. The only exceptions are the minimal $\text{D}(k,1)$ with $k\geq3$, with eight, as in \cite{Kawahigashi:2002px,Kawahigashi:2003gi,Benedetti:2024utz}, and the supersymmetric minimal models $\text{D}(k,2)$ with $k\geq3$, with twelve. For Ising $\text{D}(1,1)$ and tricritical Ising  $\text{D}(2,1)$, we find three and five submodels respectively. We provide a complete list and characterization in appendix \ref{app-gko}, and display the embeddings for generic $k$ and $l$ odd in Fig.\ref{figure-gko-letter}.

Using this classification and our algebraic selection rule, we recover the integrable RG flows triggered by perturbing D($k,l$) by the operator $\phi_{(0,0,2)}$ with left conformal dimension $
h_{(0,0,2)}=\frac{k+l}{k+l+2}$, namely  \cite{Zamolodchikov:1991vg} 
\bea
    \text{D}(k,l) &\xrightarrow{\phi_{(0,0,2)}} & \text{D}(k-l,l) \label{gkoflow1}\,,\\
    \text{D}(k,l) &\xrightarrow{\phi_{(0,0,2)}}& \text{D}(k,l-k) \label{gkoflow2}\,,
\eea
The smallest closed model containing $\phi_{(0,0,2)}$ is  
\[
\mathcal{T}_{UV}[\phi_{(0,0,2)}]\equiv\{\phi_{(0,0,\gamma)}  \hbox{ for all }\gamma \hbox{ allowed} \}\,.
\]
The global index $\mu_{\mathcal{T}}$ can be found in appendix \ref{app-gko} and it suggests $\mathfrak{su}(2)_k^{\text{even}} \times \mathfrak{su}(2)_l$ or $\mathfrak{su}(2)_k \times \mathfrak{su}(2)_l^{\text{even}}$ a candidate DHR category \footnote{In the case where both $k$ and $l$ are odd, the two categories coincide.}. This can be further refined, for example for $l$ odd, the DHR category is $\mathfrak{su}(2)_k \times \mathfrak{su}(2)_l^{\text{even}}$ and has $(k+1)(l+1)/2$ elements which can be identified with the classes of fields  $J_{\alpha,\beta}\equiv\{\phi_{(\alpha,\beta,\gamma)}\text{ for }\gamma\text{  allowed}\}$ with $\alpha=0,1,\dots k$ and $\beta=0,2,\dots l-1$. The fusion of these classes readily matches the fusion of the DHR category.

From the symmetry standpoint, $\mathcal{T}_{UV}[\phi_{(0,0,2)}]$ is characterized as the set of fields invariant under a closed fusion ring $\mathfrak{F}_{UV}[\phi_{(0,0,2)}]$ of  TDLs. Indeed, we compute that
\be
\mathfrak{F}_{UV}[\phi_{(0,0,2)}]=\{\mathcal{L}_{(\alpha,\beta,0)} \text{ for all }\alpha,\beta \text{ allowed} \}\,, \label{fusion-gko-uv}
\ee
with  $\mathfrak{F}_{UV}[\phi_{(0,0,2)}]$ forming the fusion ring of the  algebras $\mathfrak{su}(2)_k^{\text{even}} \times \mathfrak{su}(2)_l$ for $k$ odd, and $\mathfrak{su}(2)_k \times \mathfrak{su}(2)_l^{\text{even}}$ for $l$ odd, matching the DHR category of $\mathcal{T}_{UV}[\phi_{(0,0,2)}]$, as argued on general terms previously.

Following the algebraic selection rule, these DHR and symmetry fusion rings should be preserved by the RG flow, together with the structure of inclusions. When the IR is $\text{D}(k-l,l)$ in (\ref{gkoflow1}) the IR fields and TDL are
\begin{equation}\begin{split} 
\mathcal{T}_{IR}[\phi^{IR}_{(0,2,0)}] &\equiv\{\phi^{IR}_{(0,\beta,0)} \text{ for all }\beta \text{ allowed}\}\,,\\
\mathfrak{F}_{IR}[\phi^{IR}_{(0,2,0)}]&=\{\mathcal{L}^{IR}_{(\alpha,0,\gamma)} \text{ for all }\alpha,\gamma \text{ allowed}\}\,.
\end{split}\end{equation}
If the IR is $\text{D}(k,l-2)$, the categories are reproduced by
\begin{equation}\begin{split} 
\mathcal{T}_{IR}[\phi^{IR}_{(2,0,0)}] &\equiv \{\phi^{IR}_{(\alpha,0,0)} \text{for all }\alpha \text{ allowed}\}\,,\\
\mathfrak{F}_{IR}[\phi^{IR}_{(2,0,0)}]&=\{\mathcal{L}^{IR}_{(0,\beta,\gamma)} \text{ for all }\alpha,\gamma \text{ allowed}\}\,.
\end{split}\end{equation}
A pictorial representation of the algebraic selection rule can be found in appendix \ref{app-gko}. Finally, we find no new RG flows starting at the coset \eqref{coset-gko}.

\textbf{RG flows from $\mathbb{Z}_k$ parafermions:} 
We now consider the $\mathbb{Z}_k$-invariant parafermion models  defined as the coset 
\be 
\text{PF}(k)= \frac{\mathfrak{su}(2)_k}{\mathfrak{u}(1)_k}\,.\label{coset-parafermions}
\ee 
These  RCFT$_2$ contain fractional-spin operators generalizing the Majorana fermion of the Ising model, the $k$\,=\,2 case \cite{Fateev1}. Their central charge is
\be 
c_{\text{PF}(k)}=\frac{2k-2}{k+2}\,,\label{charge-parafermions}
\ee
with $\text{PF}(3)$ the three-state Potts model   \cite{Potts1,Potts2,Potts3}, and $\text{PF}(4)$ a $c$\,=\,1 orbifold theory \cite{Ginsparg:1988ui,Dijkgraaf:1987vp}. 
The chiral fields are labeled by a duo $(\lambda,\mu)$ with \cite{DiFrancesco:1997nk}
\be 
0\leq \lambda\leq k\,, \quad -k+1\leq\mu\leq k\,, \quad \, \lambda-\mu=0 \,\,\text{mod }2\,, \label{labels-parafermions}
\ee
identified by $({\lambda,\mu})\leftrightarrow({k-\lambda,k+\mu})$ and $({\lambda,\mu})\leftrightarrow ({k-\lambda,-k+\mu})$. The modular-invariant partition functions were found in \cite{Qiu1,Qiu2,Ravanini}. We again focus on diagonal theories, with fusion obtained from (\ref{verlinde-fusion}) using $S^{\mathfrak{su}(2)_k}_{\lambda\lambda'}\overline{S}^{\mathfrak{u}(1)_k}_{\mu\mu'}$, with $ S_{\mu'\mu'}^{\mathfrak{u}(1)_k}\equiv e^{-\frac{i\pi\mu\mu'}{k}}/\sqrt{2k}$,
corresponding to the compactified free boson at radius $\sqrt{2k}$. 

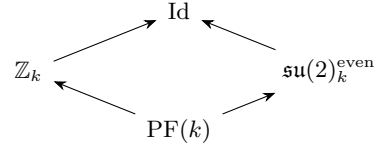
\begin{figure}[h]
\centering
\begin{tikzpicture}[
    node distance=2.2cm and 3.0cm,
    every node/.style={font=\small},
    -/.style={->, thick}
]
\node (Id)   at (-4,0.8) {$\text{Id}$};
\node (Z) at ( -6,0) {$\mathbb{Z}_k$};
\node (su) at ( -2,0) {$\mathfrak{su}(2)_k^{\text{even}}$};
\node (PF) at (-4,-0.8) {$\text{PF}(k)$};
%------------------------------------------------
\draw[arr] (Z) -- (Id);
\draw[arr] (su) -- (Id);
\draw[arr] (PF) -- (su);
\draw[arr] (PF) -- (Z);
\end{tikzpicture}
\vspace{-.2cm}
\caption{Classification of submodels of the Parafermion diagonal modular invariant \eqref{coset-parafermions} for $k$ prime. Each is denoted by their DHR category.}
\label{figure-parafermion-letter}
\end{figure}

Using our method, we find $\text{PF}(k)$ has $2S$ submodels, with $S$ the number of divisors of $k$. The ring $\mathfrak{F}$ is invertible in half of the submodels and otherwise non-invertible. See appendix \ref{app-para} for a detailed description; the inclusions for prime $k$ are given in Fig. \ref{figure-parafermion-letter}.
\vspace{-.2cm}

From these results we can understand the RG flow to a minimal model described in \cite{Fateev3}. This flow
\be 
\text{PF}(k) \xrightarrow{\phi_{(0,2)}} \text{D}(k-1,1)\:, \label{paraflow1}
\ee
is triggered by the parafermion field $\phi_{(0,2)}$ with dimension $h_{(0,2)}=1-\frac{1}{k}$. The smallest submodel with this field is
\be 
\mathcal{T}_{UV}[\phi_{(0,2)}]\equiv\{\phi_{(0,\mu)}\text{ for all }\mu\text{ allowed}\}\;,\label{paratuv}
\ee
which is the neutral part under the set of TDL given by
\be 
\mathfrak{F}_{UV}[\phi_{(0,2)}]=\{\mathcal{L}_{(\lambda,0)}\text{ for all }\lambda \text{ allowed}\}\;.
\ee
These and the DHR category have the $\mathfrak{su}(2)_k^{\text{even}}$ fusion rules. In the IR minimal model, these are reproduced by 
\begin{equation}\begin{split} 
\mathcal{T}_{IR}[\phi_{(2,0,0)}]&\equiv\{\phi^{IR}_{(\alpha,\text{Mod}[\alpha,2],0)}\text{ for all }\alpha \text{ allowed}\}\;,\\
\mathfrak{F}_{IR}[\phi_{(2,0,0)}]&=\{\mathcal{L}^{IR}_{(0,0,\gamma)}\text{ for all } \gamma \text{ allowed}\}\;,
\end{split}\end{equation}
which is consistent with the results of \cite{Feverati:1995hy}. 

We note the category $\mathfrak{su}(2)_k^{\text{even}}$ is also reproduced in the IR by the minimal model $\text{D}(k,1)$ with $k\geq 4$
\begin{equation}\begin{split} 
\mathcal{T}_{IR}[\phi_{(0,0,2)}]&\equiv\{\phi^{IR}_{(0,\text{Mod}[\gamma,2],\gamma)}\text{ for all }\gamma \text{ allowed}\}\,,\\
\mathfrak{F}_{IR}[\phi_{(0,0,2)}]&=\{\mathcal{L}^{IR}_{(\alpha,0,0)}\text{ for all } \alpha \text{ allowed}\}\,.
\label{newflow}
\end{split}\end{equation}
This flow cannot be discarded by symmetry arguments, not even by the corresponding t'Hooft anomalies \cite{hooft1980naturalness, Gaiotto:2014kfa}. Indeed, this category is preserved in the known flow between minimal models \cite{Zamolodchikov:1987ti}, i.e.\  (\ref{gkoflow1}) with $l$\,=\,1.
%\be 
%\text{D}(k,1)\xrightarrow{\phi_{(0,0,2)}}\text{D}(k-1,1)\;.
%\ee
There exists a self-dual lattice model \cite{OBrien:2019fvt} containing all three CFTs for $k$\,=\,4 in its phase diagram: D(3,1) (3-state Potts, $c$\,=\,4/5), D(4,1) (tricritical 3-state Potts, $c$\,=\,6/7), and a $U(1)$-invariant orbifold of the $\mathbb{Z}_4$ parafermions ($c$\,=\,1). Nevertheless, here and in the corresponding integrable field theory \cite{Fateev2}, the flow triggered by $\phi_{(0,2)}$ obeys (\ref{paraflow1}) for both signs of the perturbation, not \eqref{newflow}. However, including other operators with the same symmetries as $\phi_{(0,2)}$ in the perturbation may yield a different IR fixed point. We thus cannot rule out a direct flow
\be 
\text{PF}(k) \longrightarrow \text{D}(k,1)\;.
\ee
%One possibility is that the flow to  $\text{D}(k,1)$ might appear with other triggering relevant field in (\ref{paratuv}), e.g. $\phi_{(k,k-2)}$ which also scales as with $h_{(0,2)}$.  
%e.g., from $\mathbb{Z}_4$ parafermions to the tricritical three-state Potts CFT. 

Finally, for $k\geq5$, another relevant perturbation of $\text{PF}(k)$ is by $\phi_{(2,0)}$, with dimension $h_{(2,0)}=2/(k+2)$. The smallest model involving it is
\begin{equation}
\begin{split}
\mathcal{T}_{UV}[\phi_{(2,0)}]&\equiv \{\phi_{(2\lambda,0)}\,,\,\,0\leq\lambda\leq [k/2]\}\:,\\
\mathfrak{F}_{UV}[\phi_{(2,0)}]&=\{\mathcal{L}_{(0,2\mu)}\,,\,\,0\leq\mu\leq k-1\}\;,
\end{split}
\end{equation}
with the DHR category (equal again to the symmetry one) built from a $\mathbb{Z}_k$ fusion ring. This flow ends at a compact scalar field at rational radius \cite{Dorey:1996he}
\be 
\text{PF}(k) \xrightarrow{\phi_{(2,0)}} \mathfrak{u}(1)_{k}\;,
\ee
with $\mathcal{T}_{IR}$ formed by vertex operators neutral under $\mathbb{Z}_k$.

\textbf{Discussion:}  In this letter we have provided a unified and potentially complete approach to selection rules. It is based on the classification of local CFT$_2$, %(i.e. those $T$-invariant but not necessarily $S$-invariant), 
and follows from standard local CFT data. We showed that the selection rules coming from the preservation of the DHR category \cite{Benedetti:2024utz}, and the ones obtained from that of the fusion ring of TDLs  \cite{Chang:2018iay}, are already contained in such classification and associated algebraic selection rule. 
In turn, our analysis makes clear that the classification of possible $Q$-systems for chiral algebras \cite{Longo:1994zza,Longo:1994xe,LongoKawaMuger,bischoff2015tensor}, is also a classification of possible non-invertible symmetries in RCFT$_2$. We applied this framework to $\mathfrak{su}(2)$ GKO cosets (\ref{coset-gko}) and $\mathbb{Z}_k$ parafermions (\ref{coset-parafermions}), classifying all possible submodels (therefore subfactors and $Q$-systems), symmetries, and DHR categories. We then used this classification to understand selection rules for RG flows triggered by perturbations in those CFT$_2$, rederiving and generalizing existing results in the literature. Moreover, our results strongly suggest that direct flows only occur when the UV and IR CFT$_2$'s share a non-trivial subsector.  

We end by highlighting that the scope of the present method goes beyond RG flows between diagonal RCFT$_2$. A first avenue for generalization concerns non-diagonal RCFT$_2$, which requires employing the operator product expansion between primary operators, instead of its much simpler fusion rules (\ref{verlinde-fusion}). Such a procedure was required in Ref. \cite{Benedetti:2024utz} to recover all $c<1$ $Q$-systems unravelled in \cite{Kawahigashi:2002px,Kawahigashi:2003gi}, and is amenable to several $c>1$ generalizations \footnote{We note that some non-diagonal models can still be formulated as diagonal by considering a different coset description, e.g., the three-state pots model which is non-diagonal for $\text{D}(3,1)$ and diagonal for $\text{PF}(3)$. In such case the present results and those of Ref. \cite{Benedetti:2024utz} are consistent.}. Another avenue for future work concerns the extension of these methods to cases including mixed IR phases. Such a phase could involve a CFT$_2$ along with a topological field theory (see e.g.\ \cite{Chang:2018iay}), as in the interesting case of QCD-like models \cite{gomis1,gomis2}.
Finally, it would be very interesting to apply these methods to RG flows involving irrational CFT$_2$, such as the ones appearing between string theory orbifold geometries \cite{Yang:2005iua,Genenberg:2006de}.

{\bf Acknowledgements.}  We are grateful to Horacio Casini, Atish Dabholkar, Nicola Dondi, and Diego Hofman for insightful discussions. This work was initiated at the workshop, \textit{Program on Anomalies, Topology and Quantum Information in Field Theory and Condensed Matter Physics} at ICTP-SAIFR, and we thank the organizers and participants for the stimulating meeting. V.B. acknowledges the support of a RFA Fellowship from the Abdus Salam International Centre for Theoretical Physics (ICTP), Trieste, Italy, and by INFN Iniziativa Specifica ST$\&$FI. The work of J.M is supported by a Ramón y Cajal fellowship from the Spanish Ministry of Science. In the first stages of this project, the work of J.M was supported by Conicet, Argentina. The work of P.F.\ was supported in part by the UK Engineering and Physical Sciences Research Council grant EP/X030881/1.

\onecolumngrid  \vspace{0.2cm} 
%\begin{center}  
  \vspace{-0.5cm} 
%\end{center} 
\appendix

\section{Diagonal GKO cosets (sub)models}\label{app-gko}
In this section we provide a description of all possible (sub)models of the 
diagonal $\mathfrak{su}(2)$  Goddard-Kent-Olive (GKO) coset model \cite{Gko1,Gko2}.The chiral representations of the coset are labeled by $(\alpha,\beta,\gamma)$ with each one of them corresponding to one of $\mathfrak{su}(2)$ and $\gamma$ included in the tensor product $\alpha \otimes \beta$ modulo additions of the adjoint in the WZW of the numerator.  To describe the field spectrum allowed in the diagonal modular theory, we use the fields $\phi_{(\alpha,\beta,\gamma)}$ defined by taking equal holomorphic and antiholomorphic chiral representations with $(\alpha,\beta,\gamma)$.  We can obtain a unique description by taking the labels in the $\alpha=0,1\dots,k$, $\beta=0,1\dots,l$, and $\gamma=0,1\dots,k+l$ and implemnting the tensor product condition that includes $\alpha+\beta+\gamma= 0 \,\text{Mod }2$. We also need to take only one representative under the identification $(\alpha,\beta,\gamma)\leftrightarrow (k-\alpha,l-\beta,k+l-\gamma)$ of labels.  In the diagonal coset construction, the infinite irreducible Virasoro representations can be joined into a finite number of reducible representations \cite{Kmq,Ravaninigko1,Ravaninigko2,Bagger:1987kv}. Therefore, the corresponding characters can be identified with the branching functions of the obtained from the $\mathfrak{su}(2)_k$ characters. and one can check that these are consistent with the criterion above. 

As a further cross-check of these conventions, we note that the case $k=m-2$ and $l=1$ recovers the unitary minimal model series. This means that the model $\text{D}(m-2,1)$ is the minimal model with central charge $1-{6}/{m(m+1)}$, where the usual minimal model Kac labeling is recovered by $r=\alpha+1$, $s=\gamma+1$ and the condition $\beta=\text{Mod}[\alpha+\gamma,2]$ for $\text{Mod}[\alpha,2]=0$ if $\alpha$ is even and $\text{Mod}[\alpha,2]=1$ $\alpha$ is odd.  From this dictionary, we recover the minimal model characters, scaling dimensions, $S$-matrix, and fusion rules from the formulas in the main text.

\begin{table}[h]
    \centering
    \renewcommand{\arraystretch}{1.4}
\begin{tabular}{ccccc}
\cline{1-5}
\multicolumn{1}{|c|}{}  & \multicolumn{1}{c|}{Fields included}  & \multicolumn{1}{c|}{Verlinde lines} & \multicolumn{1}{c|}{DHR Category}   & \multicolumn{1}{c|}{Global Index ($\beta$)} \\ \cline{1-5}
\multicolumn{1}{|c|}{$1$}           & \multicolumn{1}{c|}{$\phi_{(\alpha,\beta,\gamma)}$ for all $\alpha$, $\beta$ and $\gamma$  }& \multicolumn{1}{c|}{$\mathcal{L}_{(0,0,0)}$ } & \multicolumn{1}{c|}{Id}     & \multicolumn{1}{c|}{$1$}               \\ \cline{1-5}
\multicolumn{1}{|c|}{$2$}           & \multicolumn{1}{c|}{$\phi_{(\alpha,\beta,\gamma)}$ for $\alpha$, $\beta$ and $\gamma$ even  }& \multicolumn{1}{c|}{$\mathcal{L}_{(0,0,0)}$, $\mathcal{L}_{(p_k,p_l,p_{k+l})}$ } & \multicolumn{1}{c|}{$\mathbb{Z}_2$}     & \multicolumn{1}{c|}{$4$}               \\ \cline{1-5}
\multicolumn{1}{|c|}{$3$}           & \multicolumn{1}{c|}{$\phi_{(0,\beta,\gamma)}$ for all $\beta$ and $\gamma$  }& \multicolumn{1}{c|}{$\mathcal{L}_{(\alpha,0,0)}$ for all $\alpha$  } & \multicolumn{1}{c|}{$\mathfrak{su}(2)_k^{\text{even}}$}     & \multicolumn{1}{c|}{$\hat{\mu}(k)/4$}               \\ \cline{1-5}
\multicolumn{1}{|c|}{$4$}           & \multicolumn{1}{c|}{$\phi_{(\alpha,0,\gamma)}$ for all $\alpha$ and $\gamma$  }& \multicolumn{1}{c|}{$\mathcal{L}_{(0,\beta,0)}$ for all $\beta$  } & \multicolumn{1}{c|}{$\mathfrak{su}(2)_l^{\text{even}}$}     & \multicolumn{1}{c|}{$\hat{\mu}(l)/4$}               \\ \cline{1-5}
\multicolumn{1}{|c|}{$5$}           & \multicolumn{1}{c|}{$\phi_{(\alpha,\beta,0)}$ for all $\alpha$ and $\beta$   }& \multicolumn{1}{c|}{ $\mathcal{L}_{(0,0,\gamma)}$ for all $\gamma$  } & \multicolumn{1}{c|}{$\mathfrak{su}(2)_{k+l}^{\text{even}}$}     & \multicolumn{1}{c|}{$\hat{\mu}(k+l)/4$}               \\ \cline{1-5}
\multicolumn{1}{|c|}{$6$}           & \multicolumn{1}{c|}{$\phi_{(0,\beta,\gamma)}$ for all $\beta$ and $\gamma$ even  }& \multicolumn{1}{c|}{$\mathcal{L}_{(\alpha,0,0)}$, $\mathcal{L}_{(\alpha,0,k+l)}$ for all $\alpha$  } & \multicolumn{1}{c|}{$\mathfrak{su}(2)_k$}     & \multicolumn{1}{c|}{$\hat{\mu}(k)$}               \\ \cline{1-5}
\multicolumn{1}{|c|}{$7$}           & \multicolumn{1}{c|}{$\phi_{(\alpha,0,\gamma)}$ for all $\alpha$ and $\gamma$ even  }& \multicolumn{1}{c|}{$\mathcal{L}_{(0,\beta,0)}$, $\mathcal{L}_{(0,\beta,k+l)}$ for all $\alpha$  } & \multicolumn{1}{c|}{$\mathfrak{su}(2)_l$}     & \multicolumn{1}{c|}{$\hat{\mu}(l)/4$}               \\ \cline{1-5}
\multicolumn{1}{|c|}{$8$}           & \multicolumn{1}{c|}{$\phi_{(\alpha,\beta,0)}$ for all $\alpha$ and $\beta$ even  }& \multicolumn{1}{c|}{$\mathcal{L}_{(0,0,\gamma)}$, $\mathcal{L}_{(0,l,\gamma)}$ for all $\gamma$  } & \multicolumn{1}{c|}{$\mathfrak{su}(2)_{k+l}$}     & \multicolumn{1}{c|}{$\hat{\mu}(k+l)$}               \\ \cline{1-5}
\multicolumn{1}{|c|}{$9$}           & \multicolumn{1}{c|}{$\phi_{(0,0,\gamma)}$,$\phi_{(0,l,\gamma)}$ for all $\gamma$  }& \multicolumn{1}{c|}{$\mathcal{L}_{(\alpha,\beta,0)}$ for all $\alpha$ and $\beta$ even  } & \multicolumn{1}{c|}{$\mathfrak{su}(2)_{k}^\text{even}\times \mathfrak{su}(2)_{l}^\text{even}$}     & \multicolumn{1}{c|}{$\hat{\mu}(k)\hat{\mu}(l)/16$}               \\ \cline{1-5}
\multicolumn{1}{|c|}{$10$}           & \multicolumn{1}{c|}{$\phi_{(0,\beta,0)}$,$\phi_{(0,\beta,k+l)}$ for all $\beta$  }& \multicolumn{1}{c|}{$\mathcal{L}_{(\alpha,0,\gamma)}$ for all $\alpha$ and $\gamma$ even  } & \multicolumn{1}{c|}{$\mathfrak{su}(2)_{k+l}^\text{even}\times \mathfrak{su}(2)_{k}^\text{even}$}     & \multicolumn{1}{c|}{$\hat{\mu}(k+l)\hat{\mu}(k)/16$}               \\ \cline{1-5}
\multicolumn{1}{|c|}{$11$}           & \multicolumn{1}{c|}{$\phi_{(\alpha,0,0)}$,$\phi_{(\alpha,0,k+l)}$ for all $\alpha$  }& \multicolumn{1}{c|}{$\mathcal{L}_{(0,\beta,\gamma)}$ for all $\beta$ and $\gamma$ even  } & \multicolumn{1}{c|}{$\mathfrak{su}(2)_{k+l}^\text{even}\times \mathfrak{su}(2)_{l}^\text{even}$}     & \multicolumn{1}{c|}{$\hat{\mu}(k+l)\hat{\mu}(l)/16$}               \\ \cline{1-5}
\multicolumn{1}{|c|}{$12$}           & \multicolumn{1}{c|}{$\phi_{(0,0,\gamma)}$ for all $\gamma$  }& \multicolumn{1}{c|}{$\mathcal{L}_{(\alpha,\beta,0)}$ for all $\alpha$ and $\beta$   } & \multicolumn{1}{c|}{\parbox{4.4cm}{\vspace{2pt}
$\mathfrak{su}(2)_{k}^\text{even}\times \mathfrak{su}(2)_{l} \,\,(k\text{ odd}) $ \\ or $\mathfrak{su}(2)_{k}\times \mathfrak{su}(2)_{l}^\text{even}\,\,(l\text{ odd}) $ \vspace{2pt}}}    & \multicolumn{1}{c|}{$\hat{\mu}(k)\hat{\mu}(l)/4$}               \\ \cline{1-5}
\multicolumn{1}{|c|}{$13$}           & \multicolumn{1}{c|}{$\phi_{(0,\beta,0)}$ for all $\beta$  }& \multicolumn{1}{c|}{$\mathcal{L}_{(\alpha,0,\gamma)}$ for all $\alpha$ and $\gamma$   } & \multicolumn{1}{c|}{\parbox{4.4cm}{\vspace{2pt}
$\mathfrak{su}(2)_{k+l}^\text{even}\times \mathfrak{su}(2)_{k}\,\,(k\text{ even}) $ \\ or $\mathfrak{su}(2)_{k+l}\times \mathfrak{su}(2)_{k}^\text{even}\,\,(k\text{ odd}) $\vspace{2pt}}}     & \multicolumn{1}{c|}{$\hat{\mu}(k+l)\hat{\mu}(k)/4$}               \\ \cline{1-5}
\multicolumn{1}{|c|}{$14$}           & \multicolumn{1}{c|}{$\phi_{(\alpha,0,0)}$ for all $\alpha$  }& \multicolumn{1}{c|}{$\mathcal{L}_{(0,\beta,\gamma)}$ for all $\beta$ and $\gamma$   } & \multicolumn{1}{c|}{\parbox{4.4cm}{\vspace{2pt}
$\mathfrak{su}(2)_{k+l}^\text{even}\times \mathfrak{su}(2)_{l}\,\,(l\text{ even})$\\ or $\mathfrak{su}(2)_{k+l}\times \mathfrak{su}(2)_{l}^\text{even}\,\,(l\text{ odd}) $ \vspace{2pt}}}      & \multicolumn{1}{c|}{$\hat{\mu}(k+l)\hat{\mu}(l)/4$}               \\ \cline{1-5}
\multicolumn{1}{|c|}{$15$}           & \multicolumn{1}{c|}{$\phi_{(0,0,0)}$, $\phi_{(p_k,p_l,p_{k+l})}$}& \multicolumn{1}{c|}{$\mathcal{L}_{(\alpha,\beta,\gamma)}$ for all $\alpha$, $\beta$ and $\gamma$ even  } & \multicolumn{1}{c|}{$(\mathfrak{su}(2)_k \times \mathfrak{su}(2)_l / \mathfrak{su}(2)_{k+l})/\mathbb{Z}_{2}$}     & \multicolumn{1}{c|}{$\hat{\mu}(k+l)\hat{\mu}(k)\hat{\mu}(l)/64$}               \\ \cline{1-5}
\multicolumn{1}{|c|}{$16$}           & \multicolumn{1}{c|}{$\phi_{(0,0,0)}$}& \multicolumn{1}{c|}{$\mathcal{L}_{(\alpha,\beta,\gamma)}$ for all $\alpha$, $\beta$ and $\gamma$  } & \multicolumn{1}{c|}{$\mathfrak{su}(2)_k \times \mathfrak{su}(2)_l / \mathfrak{su}(2)_{k+l}$}     & \multicolumn{1}{c|}{$\hat{\mu}(k+l)\hat{\mu}(k)\hat{\mu}(l)/16$}               \\ \cline{1-5}
\end{tabular}
\caption{All (sub)models of the diagonal modular invariant  $\text{D}(k,l)$  coset model with their corresponding set of commuting Verlinde lines, DHR categories and their global index for $k$ and $l$ not simultaneously even. We employ the notation $\hat{\mu}(k)$ for the index (\ref{su2-index}) and $p_k=k\big(1-(-1)^k\big)/2$.}
\label{tabla1}
\end{table}

\begin{figure}[h]  
  \centering
  \begin{minipage}{0.45\linewidth}
  \centering
      \begin{tikzpicture}[
    node distance=2.2cm and 3.0cm,
    every node/.style={font=\small},
    -/.style={->, thick}
]
%\node (topL) at (-6,6) {$k\ \text{odd}\quad l\ \text{odd}\quad k+l\ \text{even}$};
\node (Id)   at (-4,6) {$\text{Id}$};
%------------------------------------------------
\node (suke) at (-6,4.5) {$\mathfrak{su}(2)_k^{\text{even}}$};
\node (sule) at ( -2,4.5) {$\mathfrak{su}(2)_{k+l}^{\text{even}}$};
\node (Z2)  at (0,4.5) {$\mathbb{Z}_2$};
%------------------------------------------------
\node (suk)  at (-6,3) {$\mathfrak{su}(2)_k$};
\node (prod1) at (-4,3.2) {$\mathfrak{su}(2)_{k+l}^{\text{even}}$};
\node (prod12) at (-4,2.8) {$\,\,\,\,\times \mathfrak{su}(2)_k^{\text{even}}$};
\node (sul)  at ( -2,3) {$\mathfrak{su}(2)_{k+l}$};
\node (sukle) at ( 0,3) {$\mathfrak{su}(2)_{l}^{\text{even}}$};
%------------------------------------------------
\node (prod2L) at (-6,1.7) {$\mathfrak{su}(2)_{l}^{\text{even}}$};
\node (prod2L2) at (-6,1.3) {$\,\,\,\,\,\times \mathfrak{su}(2)_k^{\text{even}}$};
\node (prod2M) at (-4,1.7) {$\mathfrak{su}(2)_{k+l}^{\text{even}}$};
\node (prod2M2) at (-4,1.3) {$\times \mathfrak{su}(2)_k$};
\node (prod2R) at ( -2,1.7) {$\mathfrak{su}(2)_{k+l}^{\text{even}}$};
\node (prod2R2) at ( -2,1.3) {$\,\,\,\,\times \mathfrak{su}(2)_l^{\text{even}}$};
\node (sukl)  at ( 0,1.5) {$\mathfrak{su}(2)_{l}$};
%------------------------------------------------
\node (prod3L) at (-6,0.2) {$\mathfrak{su}(2)_{l}^{\text{even}}$};
\node (prod3L2) at (-6,-0.2) {$\,\,\,\,\times \mathfrak{su}(2)_k$};
\node (coset)  at ( -4,0) {$\text{D}(k,l)/{\mathbb{Z}_2}$};
\node (prod3R) at ( -2,0.2) {$\mathfrak{su}(2)_{k+l}^{\text{even}}$};
\node (prod3R2) at ( -2,-0.2) {$\,\,\,\,\times \mathfrak{su}(2)_l$};
%------------------------------------------------
\node (bottom) at (-4,-1.5) {$\text{D}(k,l)$};
%------------------------------------------------
\draw[arr] (suke) -- (Id);
\draw[arr] (sule) -- (Id);
\draw[arr] (Z2) -- (Id);
%------------------------------------------------
\draw[arr] (suk) -- (-6,4.35);
\draw[arr] (sul) -- (-2,4.35);
\draw[arr] (prod1) -- (sule);
\draw[arr] (prod1) -- (suke);
\draw[arr] (sukle) -- (Z2);
\draw[arr] (sul) -- (Z2);
\draw[arr] (suk) -- (Z2);
%------------------------------------------------
\draw[arr] (-6,1.9) -- (-6,2.85);
\draw[arr] (-4,1.9)-- (-4,2.65);
\draw[arr] (-2,1.9) -- (-2,2.85);
\draw[arr] (0,1.75) -- (0,2.85);
\draw[arr] (prod2L) -- (sukle);
\draw[arr] (prod2M) -- (suk);
\draw[arr] (prod2M) -- (sul);
\draw[arr] (prod2R) -- (sukle);
%------------------------------------------------
\draw[arr] (-6,0.4)  -- (-6,1.15) ;
\draw[arr] (-2,0.4) -- (-2,1.15) ;
\draw[arr] (-4,0.2)  -- (-4,1.15) ;
\draw[arr] (coset) -- (prod2L2);
\draw[arr] (coset) -- (prod2R2);
\draw[arr] (-1.4,0.25) -- (-0.2,1.2);
\draw[arr] (-5.4,0.15) -- (-0.4,1.2);
%------------------------------------------------
\draw[arr] (bottom) -- (-6,-0.35);
\draw[arr] (bottom) -- (-4,-0.15);
\draw[arr] (bottom) -- (-2,-0.35);
\end{tikzpicture}
  \end{minipage}
  \hfill
  \begin{minipage}{0.45\linewidth}
  \centering
      \begin{tikzpicture}[
    node distance=2.2cm and 3.0cm,
    every node/.style={font=\small},
    -/.style={->, thick}
]
%\node (topL) at (-6,6) {$k\ \text{odd}\quad l\ \text{odd}\quad k+l\ \text{even}$};
\node (Id)   at (-4,6) {$\text{Id}$};
%------------------------------------------------
\node (suke) at (-6,4.5) {$\mathfrak{su}(2)_l^{\text{even}}$};
\node (sule) at ( -2,4.5) {$\mathfrak{su}(2)_{k+l}^{\text{even}}$};
\node (Z2)  at (0,4.5) {$\mathbb{Z}_2$};
%------------------------------------------------
\node (suk)  at (-6,3) {$\mathfrak{su}(2)_l$};
\node (prod1) at (-4,3.2) {$\mathfrak{su}(2)_{k+l}^{\text{even}}$};
\node (prod12) at (-4,2.8) {$\,\,\,\,\times \mathfrak{su}(2)_l^{\text{even}}$};
\node (sul)  at ( -2,3) {$\mathfrak{su}(2)_{k+l}$};
\node (sukle) at ( 0,3) {$\mathfrak{su}(2)_{k}^{\text{even}}$};
%------------------------------------------------
\node (prod2L) at (-6,1.7) {$\mathfrak{su}(2)_{l}^{\text{even}}$};
\node (prod2L2) at (-6,1.3) {$\,\,\,\,\,\times \mathfrak{su}(2)_k^{\text{even}}$};
\node (prod2M) at (-4,1.7) {$\mathfrak{su}(2)_{k+l}^{\text{even}}$};
\node (prod2M2) at (-4,1.3) {$\times \mathfrak{su}(2)_l$};
\node (prod2R) at ( -2,1.7) {$\mathfrak{su}(2)_{k+l}^{\text{even}}$};
\node (prod2R2) at ( -2,1.3) {$\,\,\,\,\times \mathfrak{su}(2)_{k}^{\text{even}}$};
\node (sukl)  at ( 0,1.5) {$\mathfrak{su}(2)_{k}$};
%------------------------------------------------
\node (prod3L) at (-6,0.2) {$\mathfrak{su}(2)_{k}^{\text{even}}$};
\node (prod3L2) at (-6,-0.2) {$\,\,\,\,\times \mathfrak{su}(2)_l$};
\node (coset)  at ( -4,0) {$\text{D}(k,l)/{\mathbb{Z}_2}$};
\node (prod3R) at ( -2,0.2) {$\mathfrak{su}(2)_{k+l}^{\text{even}}$};
\node (prod3R2) at ( -2,-0.2) {$\,\,\,\,\times \mathfrak{su}(2)_k$};
%------------------------------------------------
\node (bottom) at (-4,-1.5) {$\text{D}(k,l)$};
%------------------------------------------------
\draw[arr] (suke) -- (Id);
\draw[arr] (sule) -- (Id);
\draw[arr] (Z2) -- (Id);
%------------------------------------------------
\draw[arr] (suk) -- (-6,4.35);
\draw[arr] (sul) -- (-2,4.35);
\draw[arr] (prod1) -- (sule);
\draw[arr] (prod1) -- (suke);
\draw[arr] (sukle) -- (Z2);
\draw[arr] (sul) -- (Z2);
\draw[arr] (suk) -- (Z2);
%------------------------------------------------
\draw[arr] (-6,1.9) -- (-6,2.85);
\draw[arr] (-4,1.9)-- (-4,2.65);
\draw[arr] (-2,1.9) -- (-2,2.85);
\draw[arr] (0,1.75) -- (0,2.85);
\draw[arr] (prod2L) -- (sukle);
\draw[arr] (prod2M) -- (suk);
\draw[arr] (prod2M) -- (sul);
\draw[arr] (prod2R) -- (sukle);
%------------------------------------------------
\draw[arr] (-6,0.4)  -- (-6,1.15) ;
\draw[arr] (-2,0.4) -- (-2,1.15) ;
\draw[arr] (-4,0.2)  -- (-4,1.15) ;
\draw[arr] (coset) -- (prod2L2);
\draw[arr] (coset) -- (prod2R2);
\draw[arr] (-1.4,0.25) -- (-0.2,1.2);
\draw[arr] (-5.4,0.15) -- (-0.4,1.2);
%------------------------------------------------
\draw[arr] (bottom) -- (-6,-0.35);
\draw[arr] (bottom) -- (-4,-0.15);
\draw[arr] (bottom) -- (-2,-0.35);
\end{tikzpicture}
  \end{minipage}
  \caption{Classification of submodels of the coset model D($k,l$) \eqref{coset-gko} diagonal modular invariant for $k$ odd and $l$ even (left) and for $k$ even and $l$ odd (right), with each denoted by its DHR  category. This includes the corresponding relations between $\mathfrak{su}(2)_k$, $\mathfrak{su}(2)_l$ and even projections. \label{coset-cat-supplementary}}
\end{figure}
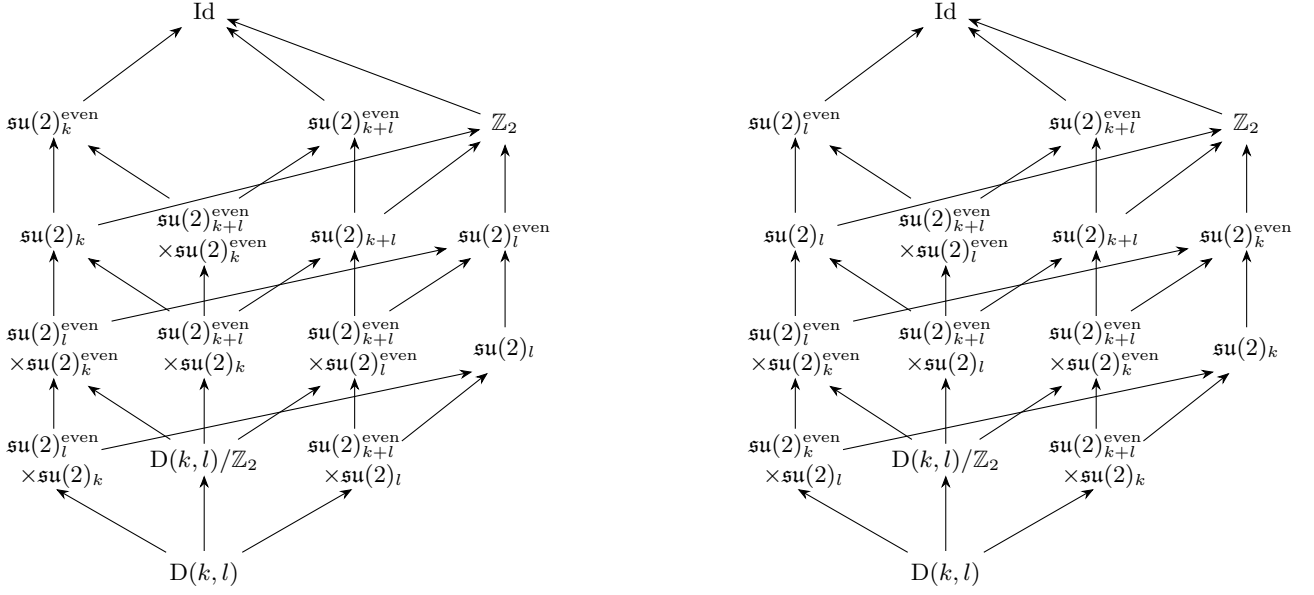

\begin{figure}[h]  
  \centering
  \begin{minipage}{0.45\linewidth}
  \centering
  \begin{tikzpicture}[
    node distance=2.2cm and 3.0cm,
    every node/.style={font=\small},
    -/.style={->, thick}
]
%\node (topL) at (-6,6) {$k\ \text{odd}\quad l\ \text{odd}\quad k+l\ \text{even}$};
\node (Id)   at (-4,6) {\textcolor{orange}{$\text{Id}$}};
%------------------------------------------------
\node (suke) at (-6,4.5) {\textcolor{orange}{$\mathfrak{su}(2)_k^{\text{even}}$}};
\node (sule) at ( -2,4.5) {\textcolor{orange}{$\mathfrak{su}(2)_l^{\text{even}}$}};
\node (Z2)  at (0,4.5) {\textcolor{orange}{$\mathbb{Z}_2$}};
%------------------------------------------------
\node (suk)  at (-6,3) {\textcolor{orange}{$\mathfrak{su}(2)_k$}};
\node (prod1) at (-4,3.2) {\textcolor{orange}{$\mathfrak{su}(2)_k^{\text{even}}$}};
\node (prod12) at (-4,2.8) {\textcolor{orange}{$\,\,\,\,\times \mathfrak{su}(2)_l^{\text{even}}$}};
\node (sul)  at ( -2,3) {\textcolor{orange}{$\mathfrak{su}(2)_l$}};
\node (sukle) at ( 0,3) {$\mathfrak{su}(2)_{k+l}^{\text{even}}$};
%------------------------------------------------
\node (prod2L) at (-6,1.7) {$\mathfrak{su}(2)_{k+l}^{\text{even}}$};
\node (prod2L2) at (-6,1.3) {$\,\,\,\,\,\times \mathfrak{su}(2)_k^{\text{even}}$};
\node (prod2M) at (-4,1.7) {\textcolor{orange}{$\mathfrak{su}(2)_k^{\text{even}}$}};
\node (prod2M2) at (-4,1.3) {\textcolor{orange}{$\times \mathfrak{su}(2)_l$}};
\node (prod2R) at ( -2,1.7) {$\mathfrak{su}(2)_{l}^{\text{even}}$};
\node (prod2R2) at ( -2,1.3) {$\,\,\,\,\times \mathfrak{su}(2)_{k+l}^{\text{even}}$};
\node (sukl)  at ( 0,1.5) {$\mathfrak{su}(2)_{k+l}$};
%------------------------------------------------
\node (prod3L) at (-6,0.2) {$\mathfrak{su}(2)_{k}^{\text{even}}$};
\node (prod3L2) at (-6,-0.2) {$\,\,\,\,\times \mathfrak{su}(2)_{k+l}$};
\node (coset)  at ( -4,0) {$\text{D}(k,l)/{\mathbb{Z}_2}$};
\node (prod3R) at ( -2,0.2) {$\mathfrak{su}(2)_{l}^{\text{even}}$};
\node (prod3R2) at ( -2,-0.2) {$\,\,\,\,\times \mathfrak{su}(2)_{k+l}$};
%------------------------------------------------
\node (bottom) at (-4,-1.5) {$\text{D}(k,l)$};
%------------------------------------------------
\draw[arr, orange] (suke) -- (Id);
\draw[arr, orange] (sule) -- (Id);
\draw[arr, orange] (Z2) -- (Id);
%------------------------------------------------
\draw[arr, orange] (suk) -- (-6,4.35);
\draw[arr, orange] (sul) -- (-2,4.35);
\draw[arr, orange] (prod1) -- (sule);
\draw[arr, orange] (prod1) -- (suke);
\draw[arr] (sukle) -- (Z2);
\draw[arr, orange] (sul) -- (Z2);
\draw[arr, orange] (suk) -- (Z2);
%------------------------------------------------
\draw[arr] (-6,1.9) -- (-6,2.85);
\draw[arr, orange] (-4,1.9)-- (-4,2.65);
\draw[arr] (-2,1.9) -- (-2,2.85);
\draw[arr] (0,1.75) -- (0,2.85);
\draw[arr] (prod2L) -- (sukle);
\draw[arr, orange] (prod2M) -- (suk);
\draw[arr, orange] (prod2M) -- (sul);
\draw[arr] (prod2R) -- (sukle);
%------------------------------------------------
\draw[arr] (-6,0.4)  -- (-6,1.15) ;
\draw[arr] (-2,0.4) -- (-2,1.15) ;
\draw[arr] (-4,0.2)  -- (-4,1.15) ;
\draw[arr] (coset) -- (prod2L2);
\draw[arr] (coset) -- (prod2R2);
\draw[arr] (-1.4,0.25) -- (-0.2,1.2);
\draw[arr] (-5.4,0.15) -- (-0.4,1.2);
%------------------------------------------------
\draw[arr] (bottom) -- (-6,-0.35);
\draw[arr] (bottom) -- (-4,-0.15);
\draw[arr] (bottom) -- (-2,-0.35);
\end{tikzpicture}
  \end{minipage}
  \hfill
  \begin{minipage}{0.45\linewidth}
  \centering
      \begin{tikzpicture}[
    node distance=2.2cm and 3.0cm,
    every node/.style={font=\small},
    -/.style={->, thick}
]
%\node (topL) at (-6,6) {$k\ \text{odd}\quad l\ \text{odd}\quad k+l\ \text{even}$};
\node (Id)   at (-4,6) {\textcolor{orange}{$\text{Id}$}};
%------------------------------------------------
\node (suke) at (-6,4.5) {\textcolor{orange}{$\mathfrak{su}(2)_l^{\text{even}}$}};
\node (sule) at ( -2,4.5) {\textcolor{orange}{$\mathfrak{su}(2)_{k}^{\text{even}}$}};
\node (Z2)  at (0,4.5) {\textcolor{orange}{$\mathbb{Z}_2$}};
%------------------------------------------------
\node (suk)  at (-6,3) {\textcolor{orange}{$\mathfrak{su}(2)_l$}};
\node (prod1) at (-4,3.2) {\textcolor{orange}{$\mathfrak{su}(2)_{k}^{\text{even}}$}};
\node (prod12) at (-4,2.8) {\textcolor{orange}{$\,\,\,\,\times \mathfrak{su}(2)_l^{\text{even}}$}};
\node (sul)  at ( -2,3) {\textcolor{orange}{$\mathfrak{su}(2)_{k}$}};
\node (sukle) at ( 0,3) {$\mathfrak{su}(2)_{k-l}^{\text{even}}$};
%------------------------------------------------
\node (prod2L) at (-6,1.7) {$\mathfrak{su}(2)_{l}^{\text{even}}$};
\node (prod2L2) at (-6,1.3) {$\,\,\,\,\,\times \mathfrak{su}(2)_{k-l}^{\text{even}}$};
\node (prod2M) at (-4,1.7) {\textcolor{orange}{$\mathfrak{su}(2)_{k}^{\text{even}}$}};
\node (prod2M2) at (-4,1.3) {\textcolor{orange}{$\times \mathfrak{su}(2)_l$}};
\node (prod2R) at ( -2,1.7) {$\mathfrak{su}(2)_{k}^{\text{even}}$};
\node (prod2R2) at ( -2,1.3) {$\,\,\,\,\times \mathfrak{su}(2)_{k-l}^{\text{even}}$};
\node (sukl)  at ( 0,1.5) {$\mathfrak{su}(2)_{k}$};
%------------------------------------------------
\node (prod3L) at (-6,0.2) {$\mathfrak{su}(2)_{k-l}^{\text{even}}$};
\node (prod3L2) at (-6,-0.2) {$\,\,\,\,\times \mathfrak{su}(2)_{l}$};
\node (coset)  at ( -4,0) {$\text{D}(k-l,l)/{\mathbb{Z}_2}$};
\node (prod3R) at ( -2,0.2) {$\mathfrak{su}(2)_{k}^{\text{even}}$};
\node (prod3R2) at ( -2,-0.2) {$\,\,\,\,\times \mathfrak{su}(2)_l$};
%------------------------------------------------
\node (bottom) at (-4,-1.5) {$\text{D}(k-l,l)$};
%------------------------------------------------
\draw[arr, orange] (suke) -- (Id);
\draw[arr, orange] (sule) -- (Id);
\draw[arr, orange] (Z2) -- (Id);
%------------------------------------------------
\draw[arr, orange] (suk) -- (-6,4.35);
\draw[arr, orange] (sul) -- (-2,4.35);
\draw[arr, orange] (prod1) -- (sule);
\draw[arr, orange] (prod1) -- (suke);
\draw[arr] (sukle) -- (Z2);
\draw[arr, orange] (sul) -- (Z2);
\draw[arr, orange] (suk) -- (Z2);
%------------------------------------------------
\draw[arr] (-6,1.9) -- (-6,2.85);
\draw[arr, orange] (-4,1.9)-- (-4,2.65);
\draw[arr] (-2,1.9) -- (-2,2.85);
\draw[arr] (0,1.75) -- (0,2.85);
\draw[arr] (prod2L) -- (sukle);
\draw[arr, orange] (prod2M) -- (suk);
\draw[arr, orange] (prod2M) -- (sul);
\draw[arr] (prod2R) -- (sukle);
%------------------------------------------------
\draw[arr] (-6,0.4)  -- (-6,1.15) ;
\draw[arr] (-2,0.4) -- (-2,1.15) ;
\draw[arr] (-4,0.2)  -- (-4,1.15) ;
\draw[arr] (coset) -- (prod2L2);
\draw[arr] (coset) -- (prod2R2);
\draw[arr] (-1.4,0.25) -- (-0.2,1.2);
\draw[arr] (-5.4,0.15) -- (-0.4,1.2);
%------------------------------------------------
\draw[arr] (bottom) -- (-6,-0.35);
\draw[arr] (bottom) -- (-4,-0.15);
\draw[arr] (bottom) -- (-2,-0.35);
\end{tikzpicture}
  \end{minipage}
  \caption{Conservation of extensions and DHR categories in the RG flow from D$(k,l)$ UV (left) and D$(k-l,l)$ IR (right) for $k$ and $l$ odd. }
  \label{rg-figure}
\end{figure}
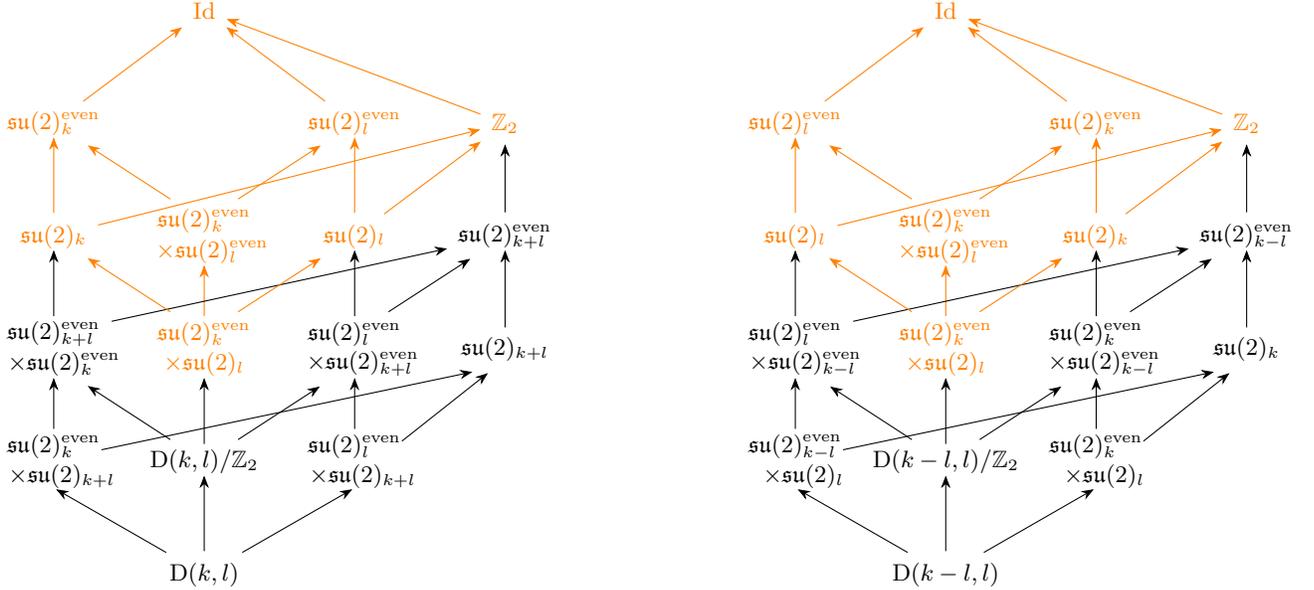

Going back to the general case for $k,l\geq 3$, searching for closed submodels as described in the main text, from the coset $S$-matrix and Verlinde formula, we find always that there are sixteen of them for each coset $\text{D}(k,l)$. We list them with their corresponding fusion ring of commuting Verlinde lines, DHR category, and global index in table \ref{tabla1} where we employ the notation $\hat{\mu}(k)$ for the global index associated with a $\mathfrak{su}(2)_k$ affine Lie algebra
\be
\hat{\mu}(k)=\Big[\sum_{\alpha=0}^{k}\big(d_\alpha^{\mathfrak{su}(2)_k}\big)^2\Big]^2=\frac{(k+2)^2}{4}\sin^{-4}\left(\frac{\pi}{k+2}\right)\,,\quad d_\alpha^{\mathfrak{su}(2)_k}=\frac{\sin\left[{\pi(\alpha+1)}/{(k+2)}\right]}{\sin\left[{\pi}/{(k+2)}\right]}\,.\label{su2-index}
\ee
which can be obtained through the $\mathfrak{su}(2)_k$ modular $S$ matrix. In table \ref{tabla1}, the cases with two categories mean that it depends on the parity of $k$ and $l$. We present them case by case in figures \ref{figure-gko-letter} and \ref{coset-cat-supplementary} according to the parity of $k$ and $l$. However, we need to take into account that there is an identification of fields  $(\alpha,\beta,\gamma)\leftrightarrow (k-\alpha,l-\beta,k+l-\gamma)$ due to the existence of a simple current that has fixed points in the case that $k$ and $l$ are both even. In these cases, the $S$ matrix needs to be resolved under this action \cite{Schellekens:1989uf,Fuchs:1995tq}, and therefore these results do not apply to these cases. For $k<3$ and/or $l<3$ this is still true, but some of the algebras overlap. For example, this happens in unitary minimal models when $k=1$ or $l=1$. In such a case, these sixteen reduce to the already known eight as in \cite{Kawahigashi:2003gi,Benedetti:2024utz}. For $k=2$ or $l=2$ in the case of $\mathcal{N}=1$ minimal models, we find twelve. For the case of Ising, where $k=1$ and $l=1$ we find three, and for Tricritical Ising, when $k=2$ and $l=1$ or vice versa we find five. Finally, using this classification, we depict in figure \ref{rg-figure} an example of the conservation of categories and extensions along the RG flow.

\newpage
\section{Parafermion coset (sub)models}\label{app-para}
A parafermion $\mathbb{Z}_k$ theory \cite{Fateev1} can be defined from the  the coset of a $\mathfrak{su}(2)_k$ WZW by its cartan $\mathfrak{u}(1)_k$ subalgebra. The spin zero fields of the diagonal modular invariant theory $\phi_{(\lambda, \mu)}$ can be uniquelly described by taking the coset labels $(\lambda, \mu)$ in the range $0\leq \lambda\leq k\,$ and  $-k+1\leq\mu\leq k$ with the constraint $\lambda-\mu=0 \,\,\text{Mod }2\,$ \cite{DiFrancesco:1997nk}. We also need to take into account the field identifications  $({\lambda,\mu})\leftrightarrow({k-\lambda,k+\mu})$ and $({\lambda,\mu})\leftrightarrow ({k-\lambda,-k+\mu})$.

\begin{figure}[h]  
  \centering
  \begin{minipage}{0.45\linewidth}
  \centering
      \begin{tikzpicture}[
    node distance=2.2cm and 3.0cm,
    every node/.style={font=\small},
    -/.style={->, thick}
]
\node (Id)   at (-4,6) {$\text{Id}$};
%------------------------------------------------
\node (su9) at (-2,4.5) {$\mathfrak{su}(2)_9^{\text{even}}$};
\node (z3) at ( -6,4.5) {$\mathbb{Z}_3$};
%------------------------------------------------
\node (p3)  at (-2,3) {$\text{PF}(9)/\mathbb{Z}_3$};
\node (z9)  at ( -6,3) {$\mathbb{Z}_9$};
%------------------------------------------------
\node (pf) at (-4,1.5) {$\text{PF}(9)$};
%------------------------------------------------
\draw[arr] (z3) -- (Id);
\draw[arr] (su9) -- (Id);
%------------------------------------------------
\draw[arr] (p3) -- (su9);
\draw[arr] (p3) -- (z3);
\draw[arr] (z9) -- (z3);
%------------------------------------------------
\draw[arr] (pf) -- (z9);
\draw[arr] (pf)-- (p3);
\end{tikzpicture}
  \end{minipage}
  \hfill
  \begin{minipage}{0.45\linewidth}
  \centering
      \begin{tikzpicture}[
    node distance=2.2cm and 3.0cm,
    every node/.style={font=\small},
    -/.style={->, thick}
]
\node (Id)   at (-4,6) {$\text{Id}$};
%------------------------------------------------
\node (z2)   at (-2,4.5) {$\mathbb{Z}_{2}$};
\node (z5)   at (-6,4.5) {$\mathbb{Z}_{5}$};
%------------------------------------------------
\node (su10) at (-2,3) {$\mathfrak{su}(2)_{10}^{\text{even}}$};
\node (z10) at ( -6,3) {$\mathbb{Z}_{10}$};
%------------------------------------------------
\node (p5)  at (-2,1.5) {$\text{PF}(10)/\mathbb{Z}_5$};
\node (p2)  at ( -6,1.5) {$\text{PF}(10)/\mathbb{Z}_2$};
%------------------------------------------------
\node (pf) at (-4,0) {$\text{PF}(10)$};
%------------------------------------------------
\draw[arr] (z2) -- (Id);
\draw[arr] (z5) -- (Id);
%------------------------------------------------
\draw[arr] (z10) -- (z2);
\draw[arr] (z10) -- (z5);
\draw[arr] (su10) -- (z2);
%------------------------------------------------
\draw[arr] (p5) -- (su10);
\draw[arr] (p2) -- (su10);
\draw[arr] (p2) -- (z10);
%------------------------------------------------
\draw[arr] (pf) -- (p2);
\draw[arr] (pf)-- (p5);
\end{tikzpicture}
  \end{minipage}
  \caption{Classification of submodels of the Parafermion diagonal modular invariant \eqref{coset-parafermions} for $k=9$ (left) and $k=10$ (right). Each is denoted by their DHR category.}
  \label{parafermions-k9k10}
\end{figure}
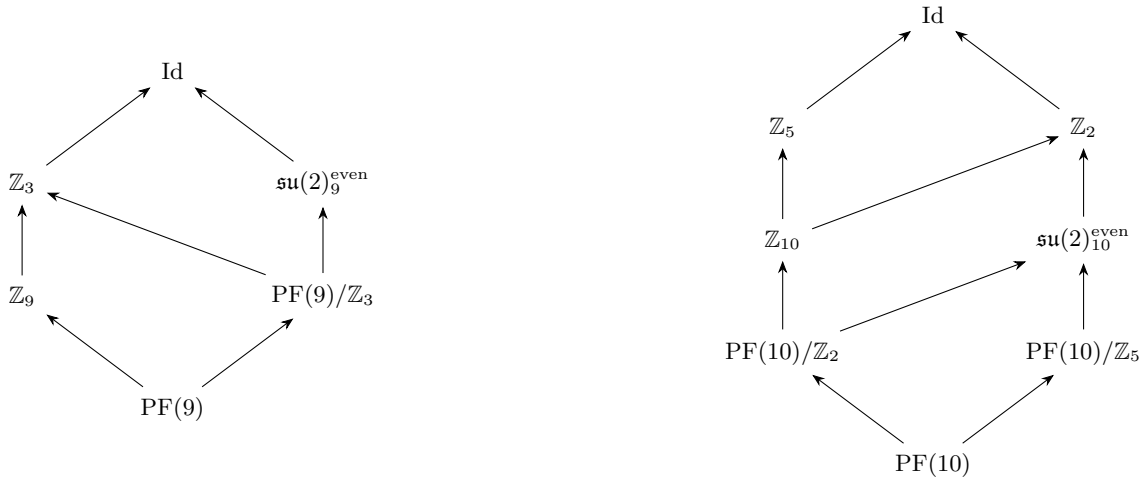

Constructing the fusion rules from the Verlinde formula, we can find all submodels. For all $k$ we find a submodel with $\mathbb{Z}_k$ superselection sectors corresponding to the fixed-point algebra of such a symmetry. We also find the same for the allowed $\mathbb{Z}_{n}$ subgroups of $\mathbb{Z}_k$. More interestingly we always find a submodel with DHR category $\mathfrak{su}(2)_k^{even}$ and the sectors of the form  $\text{PF}(k)/ \mathbb{Z}_n$ for $\mathbb{Z}_n$ for $n$ non-trivial divisors of $k$. We checked this up to $k=15$ where for $k=2$ and $k=3$ we recover the known results of Ising and the Three state potts model respectively.
These of course, come with an associated structure of inclusions as we depict for prime values of $k$ in figure \ref{figure-parafermion-letter} and $k=9,10$ in figure \ref{parafermions-k9k10}.

\begin{table}[h]
    \centering
    \renewcommand{\arraystretch}{1.4}
\begin{tabular}{ccccc}
\cline{1-5}
\multicolumn{1}{|c|}{}  & \multicolumn{1}{c|}{Fields included}  & \multicolumn{1}{c|}{Verlinde lines} & \multicolumn{1}{c|}{DHR Category}   & \multicolumn{1}{c|}{Global Index ($\mu$)} \\ \cline{1-5}
\multicolumn{1}{|c|}{$1$}           & \multicolumn{1}{c|}{$\phi_{(\lambda,\mu)}$ for all $\lambda$ and $\mu$  }& \multicolumn{1}{c|}{$\mathcal{L}_{(0,0,0)}$ } & \multicolumn{1}{c|}{Id}     & \multicolumn{1}{c|}{$1$}               \\ \cline{1-5}
\multicolumn{1}{|c|}{$2$}           & \multicolumn{1}{c|}{$\phi_{(\lambda,0)}$ for all $\lambda$   }& \multicolumn{1}{c|}{$\mathcal{L}_{(0,\mu)}$ for all $\mu$ } & \multicolumn{1}{c|}{$\mathbb{Z}_k$}     & \multicolumn{1}{c|}{$k^2$}               \\ \cline{1-5}
\multicolumn{1}{|c|}{$3$}           & \multicolumn{1}{c|}{$\phi_{(0,\mu)}$ for all $\mu$   }& \multicolumn{1}{c|}{$\mathcal{L}_{(\lambda,0)}$ for all $\lambda$  } & \multicolumn{1}{c|}{$\mathfrak{su}(2)_k^{\text{even}}$}     & \multicolumn{1}{c|}{$\hat{\mu}(k)/4$}               \\ \cline{1-5}
\multicolumn{1}{|c|}{$4$}           & \multicolumn{1}{c|}{$\phi_{(0,0)}$}& \multicolumn{1}{c|}{$\mathcal{L}_{(\lambda,\mu)}$ for all $\lambda$ and $\mu$  } & \multicolumn{1}{c|}{$\mathfrak{su}(2)_k/ \mathfrak{u}(1)_{k}$}     & \multicolumn{1}{c|}{$\hat{\mu}(k)k^2/4$}               \\ \cline{1-5}
\end{tabular}
\caption{All (sub)models of the diagonal modular invariant  $\text{PF}(k)$  coset model with their corresponding set of commuting Verlinde lines, DHR categories and their global index for $k=9$ (left)  and $k=10$ (right). We employ the notation $\hat{\mu}(k)$ for the index (\ref{su2-index}).}
\label{tabla3}
\end{table}

\vspace{-1cm}

\vspace{-0.5cm}

\bibliography{biblio}% Produces the bibliography via BibTeX.
\bibliographystyle{apsrev4-1}
\end{document}